\documentclass[twocolumn,showpacs,aps,prb,preprintnumbers,amsmath,amssymb]{revtex4}
\usepackage{graphicx}% Include figure files
\usepackage{dcolumn}% Align table columns on decimal point
\usepackage{bm}% bold math
\usepackage{epsfig}

\begin{document}

%\preprint{APS/123-QED}

\title{
Spin-orbit coupling, edge states  and  quantum spin Hall criticality \\ 
due to Dirac fermion confinement: The case study of graphene
}

\author{Grigory Tkachov and Martina Hentschel}
\affiliation{
Max-Planck Institute for the Physics of Complex Systems, Dresden, Germany
}

\begin{abstract}
We propose a generalized Dirac fermion description for 
the electronic state of graphene terminated by a zigzag edge. 
This description admits a spin-orbit coupling  
needed to preserve time-reversal invariance of the zigzag confinement, 
otherwise, for spinless particles, showing the parity anomaly typical of 
quantum electrodynamics in (2+1) dimensions.
At a certain critical strength the spin-orbit coupling 
induces a phase transition of the quantum-spin-Hall type. 
It is manifested by a novel type of the edge states 
consisting of a Kramers' pair of counter-propagating modes 
with opposite spin orientations. 
Such edge states are capable of accumulating an integer spin 
in response to a transverse electric field in the absence of a magnetic one.
They exist without any excitation gap in the bulk, 
due to which our system stands out among other 
quantum spin Hall systems studied earlier.     
We show that at the transition the local density of states 
is discontinuous and its energy dependence reflects 
the phase diagram of the system. 
\end{abstract}


\maketitle

{\bf Introduction}.-
Massless Dirac fermions in graphene, 
an isolated two-dimensional (2D) graphite layer, 
are responsible for unconventional electronic properties 
of this material~\cite{Novoselov05,Zhang05}, 
offering new functionalities for nanoelectronic devices 
such as recently realized single-electron transistors 
in graphene quantum dots~\cite{Ponomarenko08}.
This is a typical example of a situation where Dirac fermions 
occur in a confined geometry, which brings up the rather 
general issue of the boundary effects in graphene. 
The need for their characterization is 
one of the outstanding current challenges in the field,
closely related to the problem of Klein tunneling, 
and, for this reason, having no analogues 
in conventional semiconductors and metals.

\begin{figure}[b]
\begin{center}
\epsfxsize=0.9\hsize
\epsffile{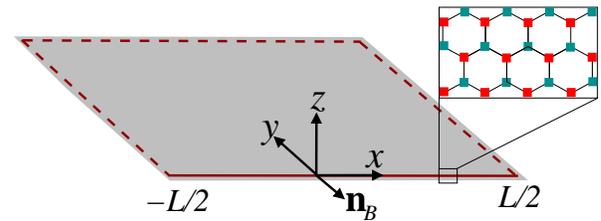}
\end{center}
\caption{
Geometry: system occupies the region $|x|\leq L/2, y\geq 0$ with a zigzag edge at $y=0$  
described by the boundary condition, Eq.~(\ref{M}), and is periodic in $x$ direction.
${\bf n}_B$ is the boundary normal.  
}
\label{Geo}
\end{figure}

In the present Letter we propose a novel spin-orbit coupling mechanism 
originating entirely from Dirac fermion confinement 
rather than being an intrinsic material property.
As a model, we consider graphene bounded by a zigzag edge 
and without spin-orbit coupling in the bulk. 
The zigzag boundary is one of the most common types 
of the honeycomb lattice termination   
(shown in Fig.~\ref{Geo}) that stands out due to its unique ability 
to support edge states, decaying in the bulk and delocalized 
along the boundary~\cite{Fujita96,Waka00, Koba05, Niimi06, Peres06, Brey06, Rycerz07, GT07, Waka07, Schomerus07, Martina07, Castro08, Akhmerov08}. 
The existence of the edge states is crucial for the spin-orbit coupling 
in our model. To demonstrate this, we first argue that for {\em spinless} 
electrons the zigzag confinement exhibits an instability toward the formation of chiral edge states. 
This would break time-reversal symmetry and drive the system into a state analogous to the integer quantum Hall phase proposed by   
Haldane~\cite{Haldane88} as a realization 
of the parity anomaly of (2+1)-dimensional quantum electrodynamics
(e.g. Ref.~\onlinecite{Semenoff84}).
	
We further show that for {\em spin-half} particles 
the requirement for the cancellation of the parity anomaly
implies the coupling between the spin and orbital 
degrees of freedom:  
Instead of the chiral edge we obtain a pair of 
counter-propagating gapless edge modes with the opposite spin orientations.  
They are related to each other by time-reversal symmetry and 
have a Kramers' degenerate spectrum, 
ensuring their robustness against time-reversal invariant perturbations.

%%%%%%%%%%%%%%%%%%%%%%%%%%%%%%%%%%%%%%%%%%%%%%%%%%%%%%%%%%%%%%%%%%%%%
%%%%%%%%%%%%%% Figure1 %%%%%%%%%%%%%%%%%%%%%%%%%%%%%%%%%%%%%%%%%%%%%%
%%%%%%%%%%%%%%%%%%%%%%%%%%%%%%%%%%%%%%%%%%%%%%%%%%%%%%%%%%%%%%%%%%%%%
%%%%%%%%%%%%%%%%%%%%%%%%%%%%%%%%%%%%%%%%%%%%%%%%%%%%%%%%%%%%%%%%%%%%%%%
%%%%%%%%%%%%%%%%%%%%%%%%%%%%%%%%%%%%%%%%%%%%%%%%%%%%%%%%%%%%%%%%%%%%%%%

Such a Kramers' doublet of gapless edge states 
is a characteristic signature of the quantum spin Hall (QSH) systems  
(e.g. Refs.~\onlinecite{Kane05,Sheng05,Bernevig06,Koenig07}).  
The recent interest in these systems is motivated by 
the principal possibility to realize a time-reversal invariant 
integer quantum Hall state in which the spin Hall conductance is quantized. 
The quantization is due to the spin accumulation ability of the edge states 
that does not require a strong magnetic field.  
The transition into the QSH state discussed in the literature~\cite{Kane05,Sheng05,Bernevig06,Koenig07}
is accompanied by opening a finite excitation gap in the 2D bulk generated, e.g. by spin-orbit interactions. 
In this respect, our case is special since zigzag-terminated graphene (as well as the unbounded system)  
is a zero-gap semiconductor, and the spin-orbit coupling at the edge does not influence the bulk electronic states.  
Nonetheless, we show that the system does exhibit a transition from 
the ordinary zero-gap semiconductor phase with spin-degenerate edge states~\cite{Fujita96,Akhmerov08}  
to the novel phase possessing a Kramers' pair of gapless spin-filtered edge modes, which can be identified as a QSH state.

The absence of the bulk energy gap makes it difficult 
to characterize the QSH state by electric transport means. 
We suggest an alternative method based on the tunneling spectroscopy 
of the local density of states (LDOS). 
In the QSH and ordinary states,
the LDOS shows drastically different energy dependences 
that cannot be continuously transformed into each other, 
implying that the two states are topologically distinct.

{\bf Discrete symmetries of the problem}.-
We start with the spinless model containing 
a pair of Weyl fermions originating from the two inequivalent valleys 
of graphene's Brillouin zone~\cite{Semenoff84}. 
They can be represented by a four-component function 
$\psi$ satisfying the Dirac equation 
$\epsilon\, \psi=-i\hbar v(\tau_z\otimes\mbox{\boldmath$\Sigma$})\mbox{\boldmath$\nabla$}\psi$
with energy $\epsilon$ measured from the Fermi level 
and momentum $-i\hbar\mbox{\boldmath$\nabla$}$ 
confined to the graphene plane, see Fig.~\ref{Geo} ($v$ is the Fermi velocity).
We consider a single edge (along the $x$-axis) that does not cause inter-valley scattering, 
which can be described by an effective boundary condition of Ref.~\onlinecite{McCann04}: 
\begin{eqnarray}
&\psi=M\psi|_{y=0},&
\nonumber\\
&
M=\cos(\Lambda\tau_z+\zeta\tau_0)\otimes\Sigma_x + 
\sin(\Lambda\tau_z+\zeta\tau_0)\otimes\Sigma_z.&
\label{M}
\end{eqnarray} 
The Pauli matrices $\Sigma_{x,z}$ act in sublattice (pseudospin) space, 
while $\tau_{z}$ and the unit matrix $\tau_0$ operate in valley space. 
Equation~(\ref{M}) contains two parameters, $\zeta$ and $\Lambda$, 
that in the limit $\zeta\to 0$, $\Lambda\to\frac{\pi}{2}$ 
yield $M=\tau_z\otimes\Sigma_z$ which is the continuum model 
for a zigzag graphene edge~\cite{Brey06,Akhmerov08}.
We intend to study the stability of the zigzag edge states 
with respect to small deviations from 
the zigzag boundary condition. 
To introduce such deviations 
we use the general formula for $M$ (\ref{M}), 
assuming small but finite parameters

\begin{equation}
\zeta\ll 1, \quad \lambda=\Lambda-\pi/2\ll 1.
\label{small}	
\end{equation}
They explicitly violate 2D parity and inversion symmetries of the system, 
respectively, allowing us to study their interplay, 
which has not been done previously~\cite{Peres06,Akhmerov08,Sasaki06}.

The 2D parity of the Dirac equation is 
defined as coordinate reflection along the edge, $x\to -x$ 
accompanied by a unitary spinor transformation~\cite{Semenoff84}, 
\begin{equation}
	\psi(x,y) \to {\cal P}\psi(-x,y), \quad 
	{\cal P}=\tau_x\otimes\Sigma_x.
	\label{P}
\end{equation}
The 2D inversion is realized by an in-plane rotation by $\pi$, yielding $x,y\to -x,-y$,  
along with the spinor transformation
\begin{equation}
	\psi(x,y) \to {\cal I}\psi(-x,-y),\quad  
       {\cal I}=\tau_0\otimes\Sigma_z.
 \label{I}
\end{equation}
The boundary condition does not share these symmetries since both 
$M^{\cal P}={\cal P} M {\cal P}^{\dagger}$ and
$M^{\cal I}={\cal I} M {\cal I}^{\dagger}$ differ from $M$ in Eq.~(\ref{M}): 
\begin{eqnarray}
&
M^{\cal P}=
\cos(\Lambda\tau_z-\zeta\tau_0)\otimes\Sigma_x + 
\sin(\Lambda\tau_z-\zeta\tau_0)\otimes\Sigma_z,
&
\label{M_P}\\
&
M^{\cal I}
=-\cos(\Lambda\tau_z+\zeta\tau_0)\otimes\Sigma_x +
\sin(\Lambda\tau_z+\zeta\tau_0)\otimes\Sigma_z.
&
\label{M_I}
\end{eqnarray} 
Setting $\zeta\to 0$ restores the parity ($M^{\cal P}\to M$), 
and if, additionally, $\lambda\to 0$ 
the inversion symmetry is recovered as well ($M^{\cal I}\to M$). 

We note that for $\zeta\not =0$ the 2D parity is 
broken simultaneously with the time-reversal ($\cal T$) 
symmetry  as the latter  is represented by the operator 
${\cal T}={\cal PC}$ [\onlinecite{Akhmerov08}] 
where ${\cal C}$ is complex conjugation. 
Despite the violation of  the $\cal T$ symmetry,
the use of Eq.~(\ref{M})  is justified  
because  $\cal T$ is only an effective symmetry of spinless 
particles (no spin degrees of freedom involved so far). 
Apart from that, the $\cal T$ symmetry breaking 
is considered as a weak perturbation with $\zeta\ll 1$. 
The small parameters  $\zeta$ and $\lambda$ compete,  
and, therefore, it is natural to carry out the stability 
analysis as a two-parametric problem. 
This analysis will eventually lead us to the  
time-reversal invariant boundary condition (\ref{M_spin}), 
which is behind the main results of this study, 
e.g. the prediction of the quantum spin Hall transition.

{\bf Spinless edge states and parity anomaly}.- 
Let us first understand the implications of the broken symmetries
for the equilibrium properties of spinless electrons 
which can be described by the LDOS,
\begin{equation}
	{\cal V}(\epsilon,{\bf r})=-(1/\pi){\rm Im Tr}\, G^R({\bf r},{\bf r}),
	\label{LDOS_def}
\end{equation}
where the trace of the retarded matrix Green's function $G^R$ is taken in $\tau\otimes\Sigma$ space. 
$G^R$ satisfies the Dirac equation,
\begin{equation}
	\epsilon\, G^R({\bf r},{\bf r}^\prime)+
i\hbar v(\tau_z\otimes\mbox{\boldmath$\Sigma$})\mbox{\boldmath$\nabla$}
G^R({\bf r},{\bf r}^\prime)=	
\delta( {\bf r}-{\bf r}^\prime ),
\label{Eq}
\end{equation}
the boundary condition (\ref{M}), and $G^R|_{y\to\infty}<\infty$.
We also assume a periodic boundary condition in the $x$ direction 
(with period $L$), modulated by a magnetic phase $\phi$:  
\begin{equation}
G^R|_{x=L/2}=G^R|_{x=-L/2}\,{\rm e}^{2\pi i\phi}. 
\label{BC2}
\end{equation}
$G^R$ can be expanded in plane waves ${\rm e}^{ik_n x}$ 
with 
\begin{equation}
	k_n=\frac{2\pi}{L}(n+\phi), \quad n \in Z (0,\pm 1,...),
\end{equation}
following from Eq.~(\ref{BC2}). 
Then, the matrix Dirac equation  
can be reduced to ordinary differential equations 
for the diagonal elements of $G^R$ which can be readily solved. 
The final result is    
\begin{eqnarray}
&&
G^R({\bf r},{\bf r}^\prime)=\sum_{\tau=\pm 1, n\in Z}
\left(\frac{\tau_0+\tau\tau_z}{2}\right)\otimes
\left( \Sigma_0+\frac{\tau\hbar v}{i\epsilon}\mbox{\boldmath$\Sigma$}
\mbox{\boldmath$\nabla$} \right)
\nonumber\\
&&
\times\left( G^{+}_{\tau k_n}(y,y^\prime)\Sigma_0 + 
G^{-}_{\tau k_n}(y,y^\prime)\Sigma_z 
\right)
\frac{{\rm e}^{ik_n(x-x^\prime)} }{L},
\label{G}
\end{eqnarray}
\begin{eqnarray}
&&
G^{+}_{\tau k_n}(y,y^\prime)=\frac{\epsilon}{2\hbar^2v^2q_n}
\left(
{\rm e}^{-q_n(y+y^\prime)} - {\rm e}^{ -q_n|y-y^\prime| }
\right)
\nonumber\\
&&
+
\frac{ q_n + k_ns_\tau }{ 2( \epsilon - \tau c_\tau\hbar vk_n ) } 
\,{\rm e}^{-q_n(y+y^\prime)},
\label{G+}
\end{eqnarray}
\begin{eqnarray}
&&
G^{-}_{\tau k_n}(y,y^\prime)=
\frac{k_n + q_ns_\tau -\tau c_\tau \epsilon/\hbar v }
{2( \epsilon - \tau c_\tau\hbar vk_n)}
\,{\rm e}^{-q_n(y+y^\prime)}.
\label{G-}
\end{eqnarray} 
Here $\tau=\pm 1$ labels the valleys and 
$q_n=\sqrt{k_n^2-\epsilon^2/\hbar^2 v^2}$.   
The edge-state spectrum follows from Eq.~(\ref{G+}) 
taken at $\epsilon\to \tau c_\tau \hbar vk_n$:
\begin{equation}
	G^{+}_{\tau k_n}(y,y^\prime)
  \approx \frac{k_ns_\tau\Theta(k_ns_\tau)}
               {\epsilon-\tau c_\tau \hbar vk_n}
  \,{\rm e}^{-k_ns_\tau (y+y^\prime)}.
  \label{G+1}
\end{equation}
It has a pole only if the unit step function $\Theta(k_ns_\tau)$ is nonzero, 
which determines the spectrum as 
\begin{equation}
\epsilon_\tau(k_n)=\tau c_\tau \hbar vk_n\quad {\rm for} \quad k_ns_\tau>0, 
\label{Spectrum}
\end{equation} 
with $c_\tau=\cos(\Lambda\tau +\zeta)$ and $s_\tau=\sin(\Lambda\tau +\zeta)$.
It is particle-hole asymmetric due to the broken discrete symmetries. 

Using Eq.~(\ref{G}) and the Poisson summation formula 
we can express the LDOS as 
\begin{eqnarray}
  &&
  {\cal V}(\epsilon,y)=-\frac{2}{\pi L}
  \sum_{\tau, n\in Z}{\rm Im}G^{+}_{\tau k_n}(y,y)=
  \label{LDOS}\\
  &&
  -\frac{1}{\pi^2}\sum_{\tau,n\in Z}{\rm e}^{-2\pi in\phi}
  \int_{-\infty}^{\infty}dk \,{\rm e}^{inkL}{\rm Im}G^{+}_{\tau k}(y,y),
\nonumber
\end{eqnarray}
where the integration over the edge ($|k|\geq |\epsilon|/\hbar v$) 
and bulk ($|k|\leq |\epsilon|/\hbar v$) 
states should be done separately. 
Accordingly, the LDOS ${\cal V}(\epsilon,y)={\cal V}_e(\epsilon,y)+{\cal V}_b(\epsilon,y)$ 
contains the edge contribution,  
\begin{eqnarray}
&&
	{\cal V}_e(\epsilon,y)=-\sum_{\tau,n\in Z}
	|2\pi\hbar v c_\tau|^{-1}\Theta(\epsilon\, \tau s_\tau c_\tau ) \times 
\label{LDOS_e}\\
&&      \times
        \exp in\left( \frac{\epsilon L }{ \hbar v c_\tau } - 2\pi\phi \right)        
	\partial_y \exp\left(
	-\frac{ 2y }{\hbar v} \left|\frac{\epsilon\, s_\tau}{c_\tau} \right| \right),
	\nonumber
\end{eqnarray}
and the bulk one,
\begin{eqnarray}
	&&
	{\cal V}_b(\epsilon,y)=\frac{|\epsilon|}{\pi^2\hbar^2v^2}\sum_{\tau,n\in Z}
	\int_0^{\pi/2}d\gamma 
	\times \label{LDOS_b}\\
	&&
	\biggl\{
	\cos(2\pi n\phi) \cos(nk_\epsilon L\cos\gamma)
	\biggl[
	1-\cos^2\gamma\times
        \nonumber\\
	&&
        \times
        \frac{s^2_\tau\cos\left( \frac{2y\epsilon}{\hbar v} \sin\gamma\right) +\tau s_\tau c_\tau\sin\gamma\sin\left( \frac{2y\epsilon}{\hbar v} \sin\gamma\right) }
	                   {s^2_\tau+c^2_\tau\sin^2\gamma}	
	\biggr]
	\nonumber\\
	&&
	+\sin(2\pi n\phi) \sin(nk_\epsilon L\cos\gamma)\sin\gamma\cos\gamma
	\times
	\nonumber\\
	&&
        \times
	\frac{\tau c_\tau\sin\gamma\cos\left(\frac{2y\epsilon}{\hbar v}\sin\gamma\right) -s_\tau\sin\left(\frac{2y\epsilon}{\hbar v}\sin\gamma\right) }
	                   {s^2_\tau+c^2_\tau\sin^2\gamma}
	\biggr\}.
	\nonumber
\end{eqnarray}
The edge LDOS (\ref{LDOS_e}) vanishes for $s_\tau=0$, 
i.e. for in-plane pseudospin orientation at the edge [see, Eq.~(\ref{M})]. 
This points to the topological origin of the edge states, 
since their existence requires a nontrivial 3D pseudospin 
structure with the out-of-plane component $\Sigma_z$.

If we now restore the symmetries, setting $\zeta,\lambda\to 0$ 
(i.e. $c_\tau\to 0$ and $s_\tau\to \tau$), 
the bulk LDOS (\ref{LDOS_b}) recovers the particle-hole symmetry, 
while the edge one, Eq.~(\ref{LDOS_e}), fails to do so because of the singularity at $c_\tau=0$. 
This anomalous asymmetry leads to a finite charge density, 
\begin{equation}
	\rho(y)=e\int_{-\infty}^0d\epsilon\, 
\left[{\cal V}(\epsilon,y)-{\cal V}(-\epsilon,y)\right]/2,
\label{rho_def}
\end{equation}
entirely localized at the edge, since the symmetric bulk LDOS cancels out.  
It can be calculated from Eq.~(\ref{LDOS_e}) as
\begin{eqnarray}
	&&
	\rho(y)=
	-\frac{e}{4L}\partial_y
	\frac{\cosh\left[
	\frac{4\pi y}{L}
	\left(\phi - \frac{ {\rm sgn}\phi }{2}\right)
	\right]\sum_\tau{\rm sgn}(\lambda +\tau\zeta )}
	 {\sinh(2\pi y/L)}
   \nonumber\\
  &&+\frac{e}{4L}\partial_y
	\frac{
	\sinh\left[
	\frac{4\pi y}{L}
	\left(\phi - \frac{ {\rm sgn}\phi }{2}\right)
	\right]
	\sum_\tau{\rm sgn}(\lambda\tau +\zeta)
	     }
	     {\sinh(2\pi y/L)},
	\label{rho}
\end{eqnarray}
where the small parameters $\zeta$ and $\lambda$ enter through the sign function, 
and the dimensionless magnetic flux is confined to the interval $0<|\phi|<1$ 
due to periodicity of Eq.~(\ref{BC2}). 
The net edge charge, $Q=L\int_0^\infty dy \rho(y)$ contains 
a flux-dependent contribution given by
\begin{eqnarray}
	Q(\phi)=Ne\left(\phi - \frac{{\rm sgn}\,\phi}{2} \right), 
	\,
  N=\frac{1}{2}\sum_{\tau=\pm 1} {\rm sgn}(\lambda\tau - \zeta).
\label{Q}
\end{eqnarray}
Remarkably, adiabatic sweeping of the flux from $0$ to $1$ 
leads to the accumulation of the integer charge 
$\Delta Q=Q(1)-Q(0)=Ne$ with $N=0,\pm 1$.
The nontrivial integers $N=\pm 1$ require $|\zeta|>|\lambda|$. 
This is the condition for the formation 
of a chiral Dirac fermion edge channel,
which follows from the edge-state spectrum, Eq.~(\ref{Spectrum}), 
linearized with respect to $\zeta$ and $\lambda$,
\begin{equation}
\epsilon_{\tau}(k_n)=-(\lambda\tau + \zeta)\hbar vk_n
\quad {\rm for} \quad k_n\tau>0.
\label{Spectrum_1}
\end{equation}    
We see that for $|\zeta|>|\lambda|$, the two graphene valleys 
prodive a {\em pair} of Weyl fermion edge states 
propagating in the same direction, i.e.  
a {\em single} chiral Dirac fermion edge channel.

The interpretation of Eq.~(\ref{Q}) is quite straightforward
if we notice that the periodicity of our system  [due to Eq.~(\ref{BC2})] 
is that of a cylinder enclosing a magnetic flux $\phi$ (in units of $ch/e$). 
Repeating Laughlin's argument~\cite{Laughlin81} 
one can identify $\Delta Q$ as the Hall charge 
accumulated in the chiral edge channel 
in response to the electric field $E_x=-(h/eL) \dot\phi$ 
generated by the adiabatically varying flux. 
On the other hand, since there is no quantizing magnetic field,
our result is close in spirit to 
Haldane's integer quantum Hall effect~\cite{Haldane88}
realizing the so-called parity anomaly~\cite{Semenoff84}. 
The anomaly consists in the odd $\phi$-dependence of $Q$,  
which is in sharp contrast to the common expectation 
that the electric charge (a scalar quantity) 
should be an even function of the magnetic flux. 
Equation~(\ref{Q}) has nevertheless normal parity.  
Indeed, the parity operation, Eq.~(\ref{P}) is 
equivalent to $\zeta, \phi  \to -\zeta, -\phi$ 
which does not affect $Q$.
Thus, the $N=\pm 1$ quantum Hall phases are characterized by 
the number of the chiral modes ($|N|=1$) and 
two possible propagation directions ($\pm$).

The parity anomaly indicates a magnetic instability 
of the continuum zigzag-edge model with respect to
the small $\cal T$ symmetry breaking perturbation.
Importantly, for $\lambda=0$ it takes only an infinitesimally small 
perturbation $\zeta$ to drive the system into 
the quantum Hall state with $N={\rm sgn}\,\zeta$.
The conventional zero-gap semiconductor state (i.e. $N=0$)
is recovered when $\zeta=0$ and $\lambda$ is finite. 
The latter could originate from 
a staggered boundary potential~\cite{Akhmerov08}.  

While the parity anomaly and its connection to the quantum Hall physics 
are of interest on their own grounds, 
in what follows we return to the $\cal T$ symmetric situation. 
The $\cal T$ invariance is recovered in the presence 
of two spin-$1/2$ fermion species experiencing   
opposite-sign quantum Hall effects.
As the spatial 2D parity remains broken, we expect  
a confinement-induced spin-orbit coupling 
with the strength controlled by $\zeta$.   
It could for instance be induced by boundary strain.
Unfortunately, we do not have the means 
to calculate the coupling strength $\zeta$ on the microscopic basis. 
Such a situation is quite common for 
the theory of spin-orbit interacting systems~\cite{Zutic04}, 
and has hardly been a serious obstacle for predicting 
new electronic properties and functionalities. 
This is also the main objective here.

{\bf Edge spin-orbit coupling and QSH transition}.-
Let us combine the pseudospinors for 
opposite spin projections $\uparrow$ and $\downarrow$ (e.g. on $z$ axis)
into a single function  $\Psi=(\psi_\uparrow,\psi_\downarrow)$
and consider the following generalization of the boundary condition, Eq.~(\ref{M}):
$\Psi={\cal M}\Psi$,
\begin{eqnarray}
{\cal M}&=&\cos(\Lambda\sigma_0\otimes\tau_z+\zeta\sigma_z\otimes\tau_0)\otimes\Sigma_x +
\nonumber\\
&+&
\sin(\Lambda\sigma_0\otimes\tau_z+\zeta\sigma_z\otimes\tau_0)\otimes\Sigma_z,
\label{M_spin}
\end{eqnarray} 
where $\sigma_0,\sigma_z$ are unit and Pauli matrices in spin space.
The presence of $\sigma_z$ makes the new boundary condition invariant 
under the time reversal operation 
${\cal T}=i\sigma_y\otimes\tau_x\otimes\Sigma_x\, {\cal C}$ 
where ${\cal C}$ is the complex conjugation. 
This also follows from the new edge spectrum 
obtained from Eq.~(\ref{Spectrum_1}) 
by $\zeta\to\zeta\sigma$ where $\sigma=\pm 1$ 
are the eigenvalues of $\sigma_z$:
\begin{equation}
\epsilon_{\tau\sigma}(k_n)=-(\lambda\tau + \zeta\sigma)\hbar vk_n,
\quad {\rm for} \quad k_n\tau>0.
\label{Spectrum_spin}
\end{equation}
In the presence of the spin-orbit term ($\propto\zeta$)
Eq.~(\ref{Spectrum_spin}) exhibits Kramers' degeneracy 
$\epsilon_{-\tau,-\sigma}(-k_n)=\epsilon_{\tau,\sigma}(k_n)$ 
as a manifestation of the time-reversal symmetry. 
Equation~(\ref{M_spin}) can be recast in the vector form 
suitable for more complicated edge geometries: 
\begin{eqnarray}
&
{\cal M}=\cos\Lambda\, \tau_0\otimes \mbox{\boldmath$\aleph$}\otimes\mbox{\boldmath$\Sigma$}+
\sin\Lambda\, \tau_z\otimes 
({\bf n}_B\times\mbox{\boldmath$\aleph$})\otimes\mbox{\boldmath$\Sigma$},
&
\nonumber\\
&
\mbox{\boldmath$\aleph$}=
\frac{1}{2}
(\mbox{\boldmath$\hat e$}-i\mbox{\boldmath$\hat z$})
{\rm e}^{i \zeta \mbox{\boldmath$\hat z\sigma$} }
+
\frac{1}{2}
(\mbox{\boldmath$\hat e$}+i\mbox{\boldmath$\hat z$})
{\rm e}^{-i \zeta \mbox{\boldmath$\hat z\sigma$} },
\label{M_spin1}
\end{eqnarray} 
where $\mbox{\boldmath$\hat e$}$ is a unit vector pointing along the edge,
$\mbox{\boldmath$\hat z$}$ is a unit vector normal to the plane, 
and the components of the vector $\mbox{\boldmath$\aleph$}$ 
are operators acting in spin space. 

%%%%%%%%%%%%%%%%%%%%%%%%%%%%%%%%%%%%%%%%%%%%%%%%%%%%%%%%%%%%%%%%%%%%%
%%%%%%%%%%%%%% Figure2 %%%%%%%%%%%%%%%%%%%%%%%%%%%%%%%%%%%%%%%%%%%%%%
%%%%%%%%%%%%%%%%%%%%%%%%%%%%%%%%%%%%%%%%%%%%%%%%%%%%%%%%%%%%%%%%%%%%%
\begin{figure}[t]
\begin{center}
\epsfxsize=0.9\hsize
\epsffile{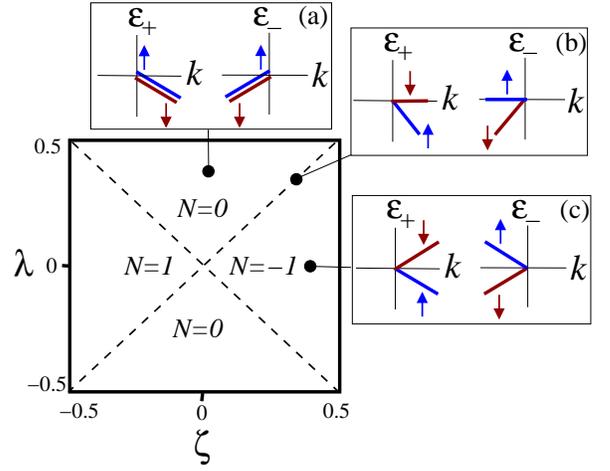}
\end{center}
\caption{
Phase diagram in space of parameters $\zeta$ and $\lambda$ 
breaking 2D parity and inversion, respectively. 
$N=0$ and $N=\pm 1$ label ordinary zero-gap semiconductor 
and quantum spin Hall states, respectively.
The critical point $\zeta=\lambda=0$ corresponds to 
the parity- and inversion-symmetric zigzag edge.  
Panels (a), (b) and (c) show the edge-state spectrum, Eq.~(\ref{Spectrum_spin}), for $\tau=\pm 1$.
}
\label{D}
\end{figure}
%%%%%%%%%%%%%%%%%%%%%%%%%%%%%%%%%%%%%%%%%%%%%%%%%%%%%%%%%%%%%%%%%%%%%%%
%%%%%%%%%%%%%%%%%%%%%%%%%%%%%%%%%%%%%%%%%%%%%%%%%%%%%%%%%%%%%%%%%%%%%%%
%%%%%%%%%%%%%%%%%%%%%%%%%%%%%%%%%%%%%%%%%%%%%%%%%%%%%%%%%%%%%%%%%%%%%%%

The two spin subsystems separately form the quantum Hall states with 
the integer factors $N_\uparrow$ and $N_\downarrow$ obtained from 
Eq.~(\ref{Q}) by substitution $\zeta\to\zeta\sigma$. 
However, there is no net Hall charge since $N_\uparrow+N_\downarrow=0$. 
Instead, from Eq.~(\ref{Q}) we find the nonzero Hall spin, 
$S(\phi)=(\hbar/2e)(Q_\uparrow(\phi)-Q_\downarrow(\phi))$,  
or, explicitly,
\begin{equation}
S(\phi)=N\hbar
\left(\phi - \frac{{\rm sgn}\,\phi}{2} \right),\quad
N=\frac{N_\uparrow-N_\downarrow}{2}.
\label{S}
\end{equation}    
Here $N(\zeta,\lambda)$ is given by Eq.~(\ref{Q}). 
It is a singular function of the symmetry breaking parameters 
which determines the phase diagram of the system shown in Fig.~\ref{D}. 
The phases with $N=0$ correspond to 
ordinary zero-gap semiconductor states with zero spin accumulation 
$\Delta S=S(1)-S(0)=0$, while the phases with $N=\pm 1$
can be identified as quantum spin Hall (QSH) states~\cite{Kane05,Sheng05,Bernevig06,Koenig07}
in which $\Delta S=\pm\hbar$. 
This identification is supported by the fact that the two valleys, 
$\tau=\pm 1$,  
provide a pair of spin-dependent edge states related by time-reversal symmetry 
[panel (c) in Fig.~\ref{D}]. 
Within such a pair the intervalley backscattering 
is forbidden unless there are time-reversal noninvariant interactions 
mixing the valleys and spin projections. 
In contrast, in the ordinary zero-gap semiconductor ($\zeta=0$ and $N=0$) 
the edge states are spin degenerate [panel (a) in Fig.~\ref{D}], and, 
therefore, spin-independent intervalley backscattering 
can lead to Anderson localization~\cite{Waka07}.
This is, for instance, the case for the edge states 
originally predicted by Fujita et al.~\cite{Fujita96} 
and their later generalizations (e.g. Ref.~\onlinecite{Akhmerov08}). 
Also, unlike QSH topological insulators~\cite{Kane05,Sheng05,Bernevig06,Koenig07},
in our case the edge states exist without any spin-orbit bulk energy gap,
since they are supported by the nontrivial pseudospin structure of the zigzag boundary. 
The novelty of this type of edge states is most prominently manifested by
the nonvanishing spin accumulation,
\begin{equation}
	S(\phi\to 0)=-N(\hbar/2){\rm sgn}\,\phi,
	\label{S0}
\end{equation}
resulting from the zero mode $n=0$ in Eq.~(\ref{LDOS_e}).

%%%%%%%%%%%%%%%%%%%%%%%%%%%%%%%%%%%%%%%%%%%%%%%%%%%%%%%%%%%%%%%%%%%%%
%%%%%%%%%%%%%% Figure3 %%%%%%%%%%%%%%%%%%%%%%%%%%%%%%%%%%%%%%%%%%%%%%
%%%%%%%%%%%%%%%%%%%%%%%%%%%%%%%%%%%%%%%%%%%%%%%%%%%%%%%%%%%%%%%%%%%%%
\begin{figure}[t!]
\begin{center}
\epsfxsize=0.95\hsize
\epsffile{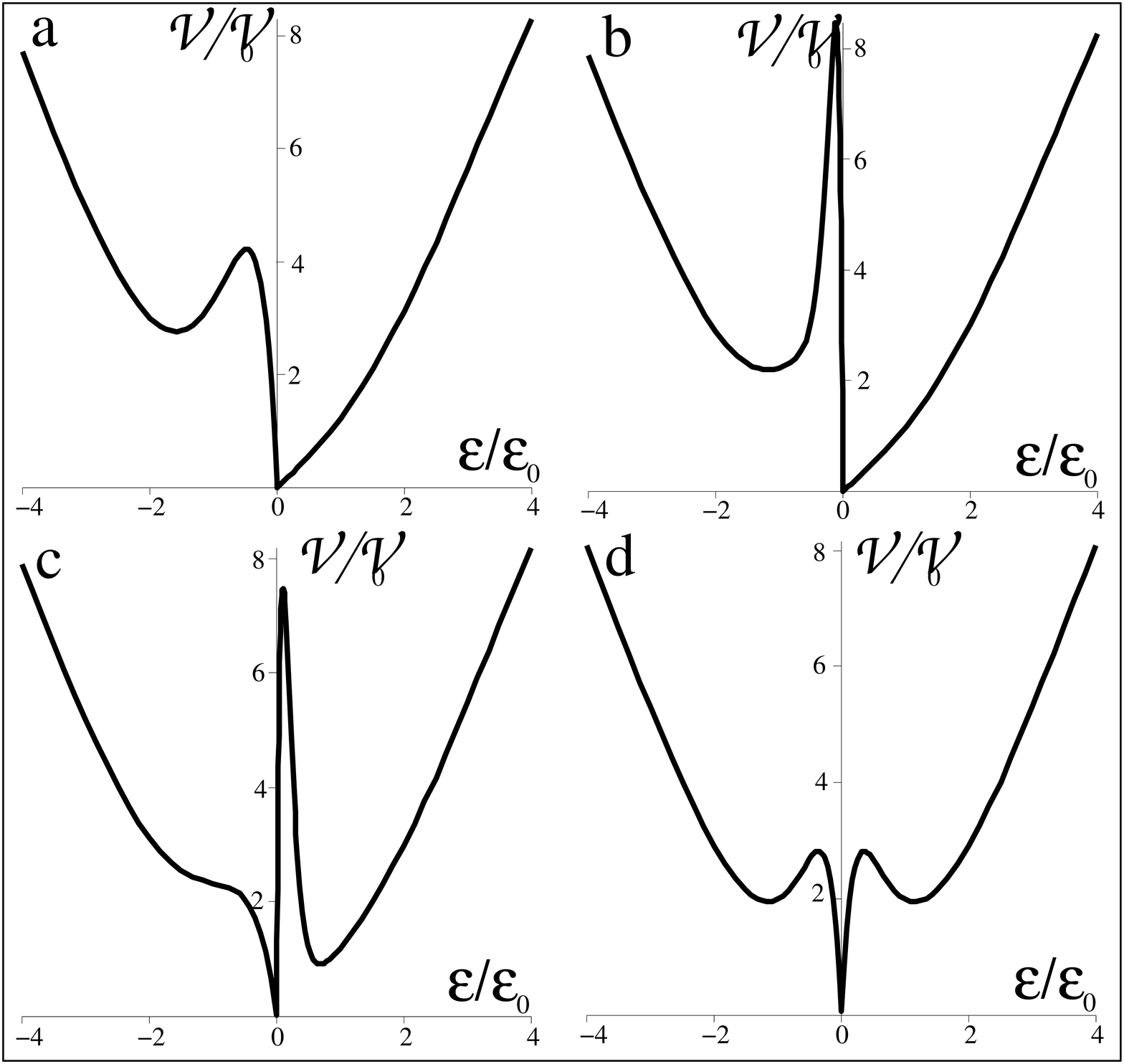}
\end{center}
\caption{
LDOS at representative points of phase diagram in Fig.~\ref{D}: 
(a) $N=0$ phase: $\zeta=0,\lambda=0.4$, 
(b) close to transition from $N=0$ side: $\zeta=0.15,\lambda=0.25$, 
(c) close to transition from $N=-1$ side: $\zeta=0.3,\lambda=0.2$, 
and (d) $N=-1$ phase: $\zeta=0.3,\lambda=0$.
We took the limit $L\to\infty$ 
in which only the $n=0$ terms in Eqs.~(\ref{LDOS_e}) and (\ref{LDOS_b}) 
contributed; $\varepsilon_0=\hbar v/2y$ and ${\cal V}_0=1/hvy$.
}
\label{LDOSfig}
\end{figure}
%%%%%%%%%%%%%%%%%%%%%%%%%%%%%%%%%%%%%%%%%%%%%%%%%%%%%%%%%%%%%%%%%%%%%%%
%%%%%%%%%%%%%%%%%%%%%%%%%%%%%%%%%%%%%%%%%%%%%%%%%%%%%%%%%%%%%%%%%%%%%%%
%%%%%%%%%%%%%%%%%%%%%%%%%%%%%%%%%%%%%%%%%%%%%%%%%%%%%%%%%%%%%%%%%%%%%%%

Because of the gapless 2D bulk, 
transport measurements are hardly suitable 
for the characterization of the $N=\pm 1$ and $N=0$ phases. 
We propose a more robust method based on 
the tunneling spectroscopy of the LDOS. 
Figure~\ref{LDOSfig} shows  
${\cal V}(\epsilon,y)=\sum_{\sigma}
\left[{\cal V}^\sigma_e(\epsilon,y)
+{\cal V}^\sigma_b(\epsilon,y)
\right]$, 
where ${\cal V}^\sigma_{e,b}(\epsilon,y)$ 
are given by Eqs.~(\ref{LDOS_e}) and (\ref{LDOS_b}) 
with $\zeta\to\zeta\sigma$. 
Panels (a) and (d) correspond to the $N=0$ and $N=-1$ phases, 
respectively, which differ by the symmetry of the energy dependence 
of ${\cal V}(\epsilon,y)$. 
The particle-hole asymmetric LDOS of the $N=0$ phase  
transforms into the symmetric LDOS of the $N=-1$ phase
through a discontinuity [Figs.~\ref{LDOSfig}(b) and (c)] 
coming from zero-energy states existing 
on the critical lines of the diagram in Fig.~\ref{D}. 
The discontinuous phase transition implies that 
the $N=\pm 1$ states are topologically distinct from 
an ordinary zero-gap semiconductor with $N=0$.

{\bf Conclusions}.- 
Within the continuum model for zigzag-terminated graphene, 
we have demonstrated the possibility of confinement-mediated 
spin-orbit coupling. The system exhibits a phase transition 
into a quantum spin Hall state in the sense that it possesses
a Kramers' doublet of spin-dependent edge states, 
accumulating an integer spin, albeit there is no true gap 
in the system's bulk as in the usual quantum spin Hall systems. 
We show that the local density of states can be used to 
distinguish the quantum spin Hall state from ordinary zero-gap 
semiconductor state of zigzag-terminated graphene.
Our findings also imply that zigzag graphene edges can be spin-active 
without interaction-induced magnetism~\cite{Fujita96,Son06,Yazyev08,Wimmer08}.

We thank H. U. Baranger, F. Guinea, M. I. Katsnelson and A. D. Mirlin 
for discussions. The work was supported by the Emmy-Noether Programme 
of the German Research Foundation (DFG).

%%%%%%%%%%%%%%%%%%%%%%%%%%%%%%%%%%%%%%%%%%%%%%%%%%%%%%%%%%%%%%%%%%%%%%%

\end{document}